# A bond counting model for accurate prediction of lattice parameter of bcc solid solution alloys

Chris Tandoc[1], Liang Qi[2], and Yong-Jie Hu[1,2]*

1Department of Material Science and Engineering, Drexel University, Philadelphia, PA, USA,
2Department of Materials Science and Engineering, University of Michigan, Ann Arbor, USA
*Corresponding authors: Yong-Jie Hu (yh593@drexel.edu)

Lattice Parameter is an important material feature in High Entropy Alloy (HEA) Design. Vegard's Law is typically used to estimate lattice parameters but is often inaccurate for metal alloys due to an inability to account for charge transfer which can affect atomic volumes. The present study used ab-initio simulation to calculate bond lengths between atoms of dissimilar elements in B2 intermetallic compounds which was then combined with a bond counting model to produce a model to estimate the lattice parameters of Refractory BCC MPEAS. The model was tested using a supercell method which modeled various random solid solution MPEAs. The proposed model produced lattice parameters with superior accuracy to Vegard's Law without the need for large DFT calculations or fitting parameters. The proposed model had a root mean squared error (RMSE) of 0.006Å which is half that of Vegard's Law (RMSE = 0.012Å).

## **Introduction**

The lattice parameter of a crystalline material is an important feature in materials design which measures the distance between unit cells in a given crystallographic direction. In cubic crystal structures such as Body Centered Cubic (BCC), there is only one lattice parameter since the unit cell has equal dimensions along all three lattice vectors. Lattice parameters can be used can be used to predict properties in single phases such as band gap in semiconductors[1]. The Lattice parameter is also important in multiphase materials due to lattice misfit between phases with implications for mechanical properties[2], thermal conductivity[3], and ionic conductivity[4]. In high entropy alloys (HEA), the lattice parameter is used to calculate a random alloy's volume mismatch parameter which in turn is an important parameter when modeling solid solution strengthening[5].

Refractory High Entropy Alloys (RHEAs), which are composed of multiple (4+) constituent elements in a body centered cubic (BCC) random solid solution, span a large composition space that is impractical to explore experimentally by trial and error. A method to predict HEA rapidly and accurately is therefore highly desirable to enable high throughput screening of the HEA composition space to optimize the HEA design process. Vegard's Law, an empirical rule which states that the lattice parameter of a solid solution can be modeled as a linear interpolation between the pure elemental lattice parameters of its constituent elements, is commonly used to estimate lattice parameters for metal alloys due to its ease of use despite widespread observation that very few alloy follow Vegard's Law[6–10]. Vegard's Law works well in ionic compounds, such as those it was developed on[11], since the ionic radii of elements generally remain constant regardless of local chemistry. Atomic volumes in metals, however, can vary depending on their local environment due to charge transfer between atoms with different electronegativities and valencies[12–14] which can explain why metal alloys rarely follow Vegard's Law.

First principles calculations based on density functional theory (DFT) are an ideal tool to capture the charge transfer effect on atomic volume in metal alloys due to its accurate description of interatomic bonding. The supercell method based on special quasi-random structures (SQS)[15] can be used to mimic the randomness of solid solution mixing. The alloy's atomic structure at ground state can be obtained by performing a relaxation calculation in which the atomic structure is relaxed by moving atoms to their equilibrium bonding distance by minimizing Hellmann-Feynman forces. The lattice parameter can be derived from the relaxed structure using the volume of the relaxed structure and the number of atoms in the supercell. Calculating lattice parameter using DFT relaxation can be computationally intensive since the cost of DFT calculations scales exponentially with the number of atoms simulated. The number of atoms needed for an SQS to adequately model a random alloy is sufficiently large to make high throughput combinatorial lattice parameter screening by DFT impractically expensive. A surrogate model to quickly estimate an alloy's lattice parameter would therefore be a valuable tool for HEA design.

DFT relaxation calculations of B2 intermetallic unit cells can be used to calculate the bond length between atoms of any two given elements ($i,j$) in a BCC random solid solution alloy due to DFT's ability to account for charge transfer effect between dissimilar atoms. The structure of the B2 unit cell is arranged such that every nearest neighbor of an atom of element $i$ will be an atom of element $j$ and vice versa. The B2 unit cell is shown in figure 1 and has the same atomic positions as a BCC unit cell with the distinction that the atoms at the corners are all one type of element while the atom in the center is of a different element. If one assumes that charge transfer primarily occurs between first nearest neighbors, then one could expect an accurate measurement of the first nearest neighbor bond length between two atoms of any two given elements. DFT relaxation of B2 unit cells require very little computational resources since they only consist of 2 atoms. Pure elemental BCC unit cells, which also consist of 2 atoms can be used to calculate the bond lengths between two atoms of the same element. Since DFT relaxation calculations for B2 and BCC are computationally light, it is possible to calculate the bond length for every possible pair combination of elements in the refractory element space being investigated. The pairs considered and their DFT calculated bond lengths can be seen in Table 1. The B2 unit cell was thus used in the present work to calculate the bond length between two atoms of elements $i$ and $j$ in a BCC random alloy.

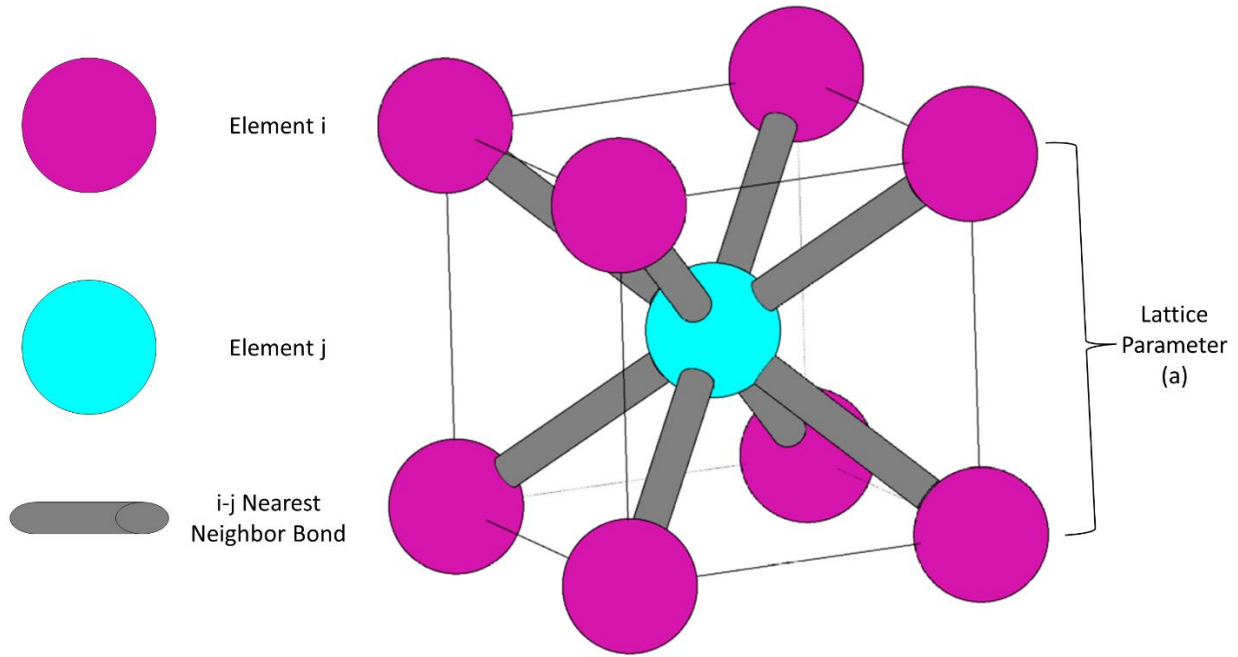

Figure 1. Schematic illustration of (a) binary B2 Unit Cell and (b) single element BCC unit cell used to model first nearest neighbor bonds.

With the DFT calculated bond lengths for elemental pairs, an average bond length in a given alloy composition can be calculated using a bond counting model which assumes that the relative frequency of a bond between any two elements in a random solid solution are equal to the product of the atomic fractions of the respective elements. The average bond length ($\bar{X}$) of a random solid solution alloy composed of $n$ principle elements can then be calculated by summing the length of an $i$-$j$ bond ($X_{i,j}$) weighted by the product of the atomic fractions of element $i$ ($c_i$) and element $j$ ($c_j$) as shown in equation 1.

$$\bar{X} = \sum_{i,j}^{n} c_i c_j X_{i,j}$$

The use of a bond counting model to predict bulk properties has been demonstrated in other materials systems[16,17] and more specifically a bond counting model based on B2 unit cell calculations has been used in our previous work to generate descriptors for BCC MPEAs[18]. Since $\bar{X}$ is half the length of the averaged BCC unit cell body diagonal, the lattice parameter can be derived using the geometric relationship between the body diagonal of a cube and its edge length: $a = \frac{2\bar{X}}{\sqrt{3}}$.

DFT calculations were performed using the projector augmented wave method (PAW)[19] with the exchange-correlation functional depicted by the general gradient approximation from Perdew, Burke, and Ernzerhof (GGA-PBE)[20] as implemented in the Vienna ab-initio simulation package (VASP) 10. The calculation parameters used were the same as those used in previous work[18] with a plane-wave basis energy cutoff of 400 eV, first-order Methfessel Paxton smearing of 0.2eV, and using k-point grids generated by VASP's automatic meshing scheme with an $R_k$ length of 30Å. Convergence criteria of $10^{-6}$ eV for electronic self-consistency cycles and $10^{-3}$ eV were used for ionic cycles.

B2 lattice parameters for the bond counting model were calculated by relaxing B2 unit cells with restrictions in place to maintain cubic symmetry such that only the volume of the unit cell was free to be relaxed. The resulting B2 lattice parameters are shown in Table 1.

B2 $i$-$j$ 1st Nearest Neighbor Bond Length (Å)

| | | Ti | Zr | Hf | V | Nb | Ta | Mo | W | Re | Ru |
|---|---|---|---|---|---|---|---|---|---|---|---|
| Element$_j$ | Ti | xxx | | | | | | | | | |
| | Zr | xxx | xxx | | | | | | | | |
| | Hf | xxx | xxx | xxx | | | | | | | |
| | V | xxx | xxx | xxx | xxx | | | | | | |
| | Nb | xxx | xxx | xxx | xxx | xxx | | | | | |
| | Ta | xxx | xxx | xxx | xxx | xxx | xxx | | | | |
| | Mo | xxx | xxx | xxx | xxx | xxx | xxx | xxx | | | |
| | W | xxx | xxx | xxx | xxx | xxx | xxx | xxx | xxx | | |
| | Re | xxx | xxx | xxx | xxx | xxx | xxx | xxx | xxx | xxx | |
| | Ru | xxx | xxx | xxx | xxx | xxx | xxx | xxx | xxx | xxx | xxx |

Table 1. Ground state 1st Nearest Neighbor Bond Length for B2 unit cells optimized by DFT relaxation of unit cell volume.

The proposed model was tested on a data set of supercells using special quasi-random structures (SQS) to approximate random chemical disorder in solid solution alloys[15]. The SQS were generated using an algorithm implemented in the Alloy Theoretic Automated Toolkit (ATAT)[22] which searches for periodic structures that most closely match the correlation functions of the resulting supercell to those of an ideally mixed solid solution. Binary, Ternary, Quaternary, and Quinary BCC solid solution alloys were modeled in this manner from a 10 element composition space comprised of group IV–VIII elements: Ti, Zr, Hf, V, Nb, Ta, Mo, W, Re, and Ru. SQS for 218 different compositions were created. The average lattice parameter ($a$) was calculated by dividing the total volume of the relaxed supercell by the number of atoms in the supercell to get the average atomic volume ($\bar{v}$) of the supercell. The lattice parameter ($a$) can be derived from $\bar{v}$ as $a = \sqrt[3]{2\bar{v}}$. The cube root of $2\bar{v}$ is used since a BCC unit cell consists of 2 atoms.

Vegard's Law and the proposed B2 bond counting model were used to make predictions for the lattice parameter of alloys modeled in the dataset. The model's results are plotted against the DFT calculated lattice parameters in figure 1a. While

Vegard's Law was fairly accurate (RMSE = 0.012 Å), the proposed model reduced the Root Mean Square Error (RMSE) by half (RMSE = 0.006 Å). The residual error of Vegard's Law and the proposed model's lattice parameter predictions were stratified by the number of principle components in the boxplots in figure 1b) and figure 1c). The boxplots show that the proposed model reduces error regardless of the number of components present in the simulated alloy. The boxplots also suggest a slight tendency for Vegard's Law predictions to overestimate lattice parameter which is evidenced by the positive mean error of non-trivial magnitude (0.005 Å, $t(220)=6.43, p<0.001$). No such bias was observed in the predictions from the proposed model (-0.0003 Å, $t(220)=-0.75, p=0.45$).

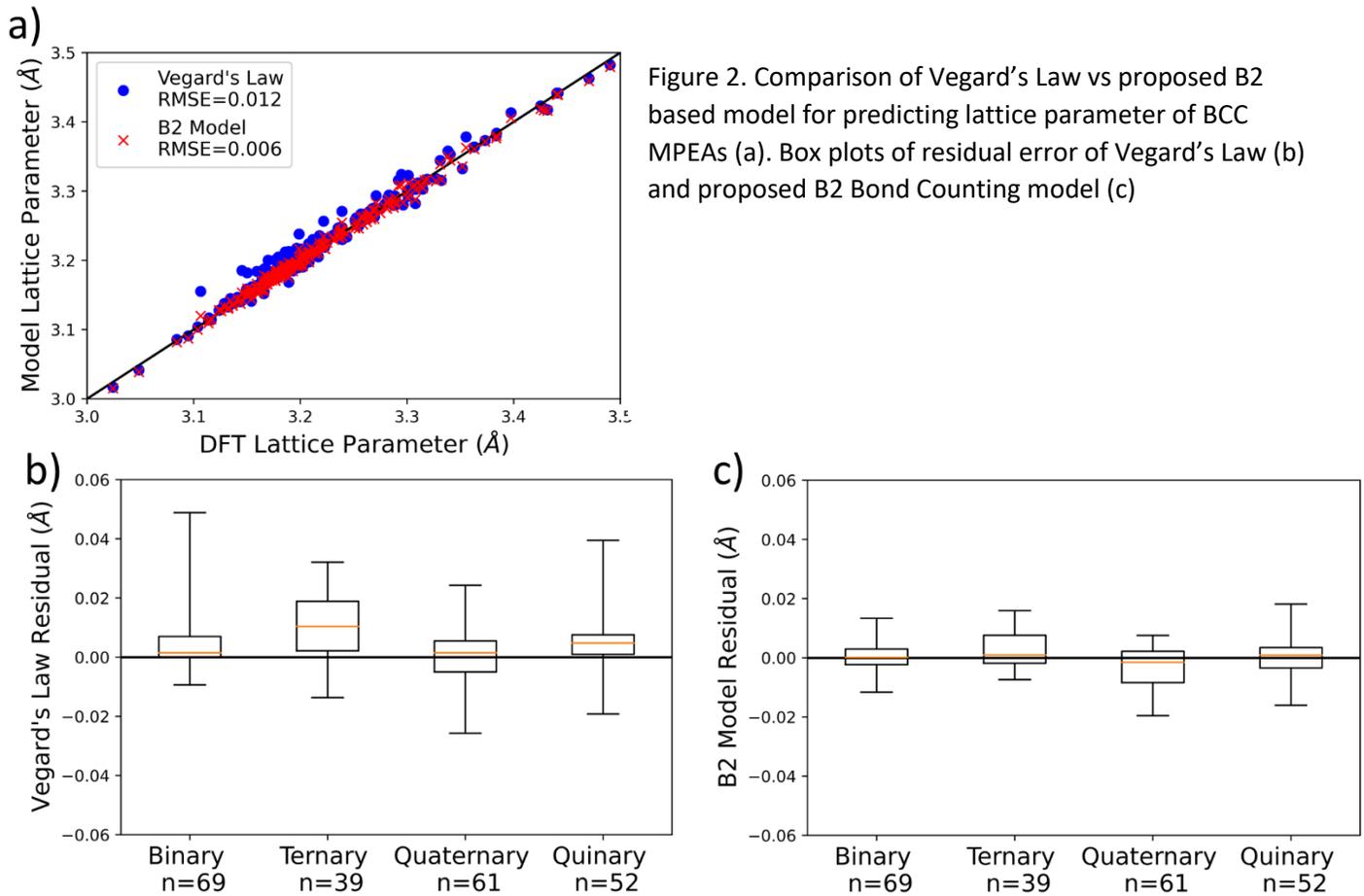

Figure 2. Comparison of Vegard's Law vs proposed B2 based model for predicting lattice parameter of BCC MPEAs (a). Box plots of residual error of Vegard's Law (b) and proposed B2 Bond Counting model (c)

The proposed model had the greatest improvement in lattice parameter prediction accuracy over Vegard's Law in the Ru-Ti and Ti-W binary series. Lattice parameters from the aforementioned compositions as predicted by the proposed model as well as the true lattice parameter calculated by DFT are plotted as a function of the binary composition in figures 2a) and 2b). Other binary systems such as Mo-Ta and Re-Ta were identified with similar trends to those seen in Ru-Ti and Ti-W and are similarly plotted in figures 2c) and 2d). A common feature shared by the aforementioned binary series is variability in electronegativity (EN) and Valence Electron Concentration (VEC) amongst the constituent elements. Indeed, the trend is seen across the majority of the data set with a strong correlation between the standard deviation of EN ($\sigma_{EN}$) and VEC ($\sigma_{VEC}$)

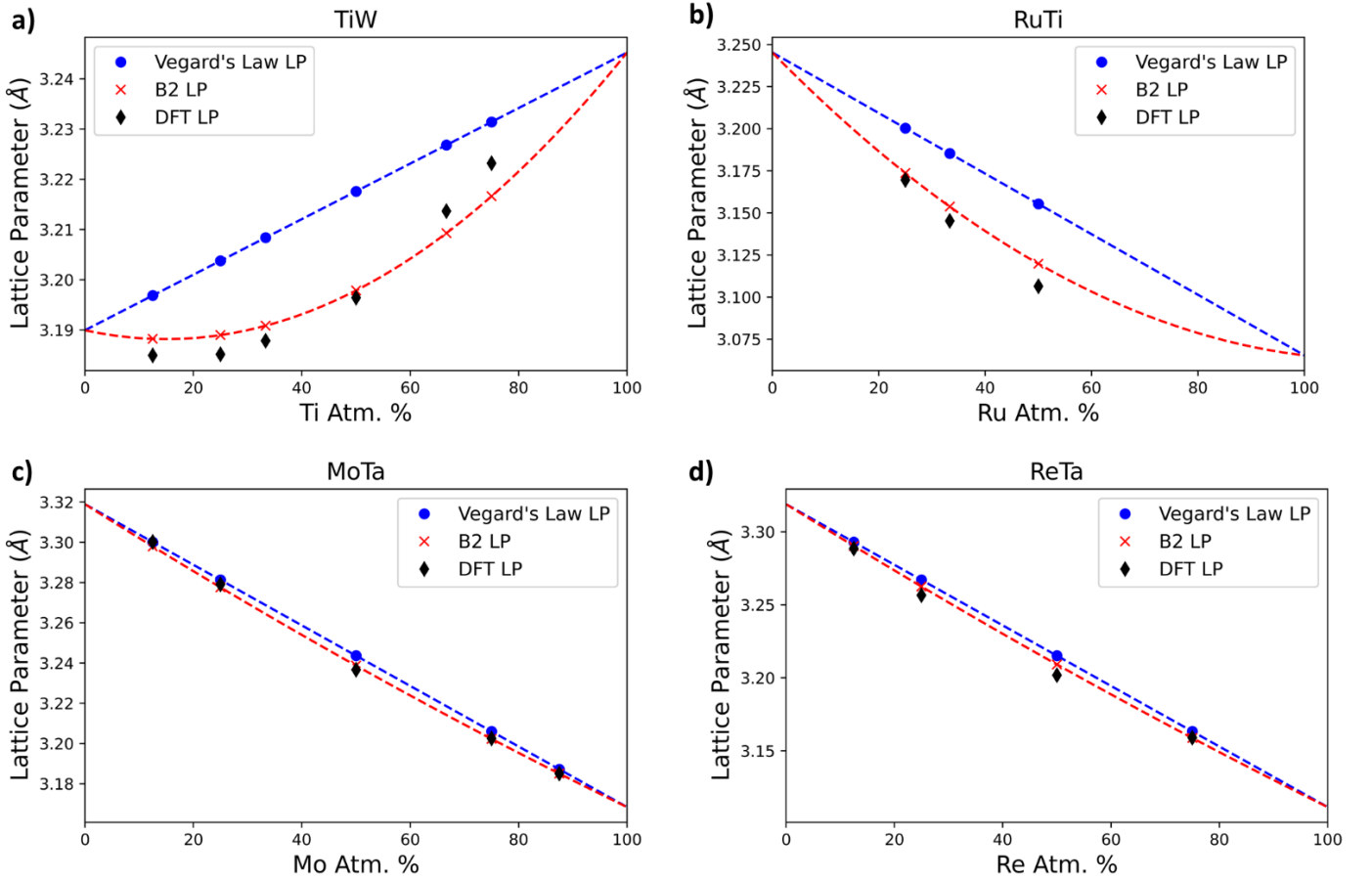

Figure 3. Plots of Vegard's Law, Proposed B2 model, and DFT calculation of lattice parameter against composition for the binary systems: Ti-W (a), Ru-Ti (b), Mo-Ta (c), Re-Ta (d).

The standard deviations $\sigma_{VEC}$ and $\sigma_{EN}$ were calculated by treating the valency, EN and frequency of constituent elements as a discrete probability distribution. The calculation used is shown in more detail in equation 2

$$\sigma_p = \sum_i^n \sqrt{(x_i - \bar{p})^2 c_i}$$

in which $p$ is either VEC or EN, $n$ is the number of principal elements, $p_i$ is the valence or EN of element $i$, $c_i$ is the concentration of element $i$, and $\bar{p}$ is the average of property $p$ calculated as the mean of a discrete probability distribution as calculated in equation 3.

$$\bar{p} = p_i c_i$$

The residual error of both Vegard's Law and the proposed model are plotted against the product of $\sigma_{VEC}$ and $\sigma_{EN}$ in figure 3a). The amount of improvement from the proposed model over Vegard's Law is plotted in figure 3b) stratified by the number of principal elements in the alloy. The reduction of atomic volume due to charge transfer has been observed in previous work on RHEAs[13,14] and is further supported here by the significant correlation between $\sigma_{VEC}$ x $\sigma_{EN}$ and the residual error of Vegard's Law correlation (r=0.76, p<10$^{-22}$). The improvement in prediction accuracy is shown in figure 3b) where it can be seen that the present work achieves the most improvement in alloys which contain more variability in VEC and EN. It can also be seen through the color coding associated with the number of principle elements that this trend is consistent regardless of the number of principle elements.

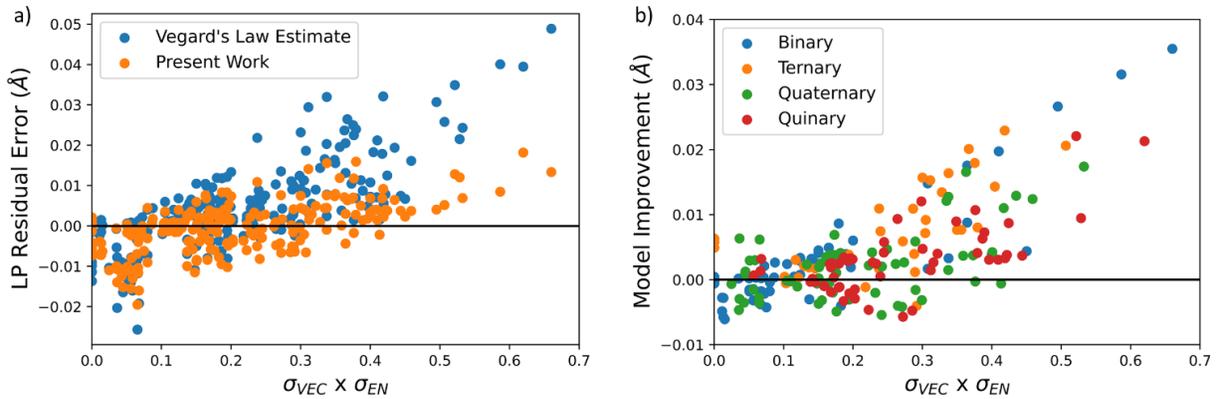

Figure 4. Exploration of the effect of variance in VEC on accuracy of Vegard's Law and the proposed model. Residual errors of the two models are plotted in a). The Improvements in the proposed model's predictions are plotted in b) with improved accuracy represented in the positive-y direction and decreased accuracy represented in the negative-y direction.

To further investigate the connection between charge transfer and atomic volume shrinkage, Bader Charges ($\rho$)[23] were calculated for one alloy SQS in which Vegard's Law significantly overestimated the lattice parameter (WTi), and one in which Vegard's Law has less error (TaMo). The results are presented in figure 5. Figures 5a and 5b are histograms of the Bader charges of atoms in TaMo and WTi respectively with tables of descriptive statistics. In these figures it can be seen that the average Bader charge for constituent elements in SQS are generally closer to the Bader charge of the same element in its corresponding B2 unit cell than pure elemental BCC. The reduced RMSE of the B2 Bader Charges show that the use of B2 unit cells to calculate bond lengths captures charge transfer in a way that Vegard's Law can not.

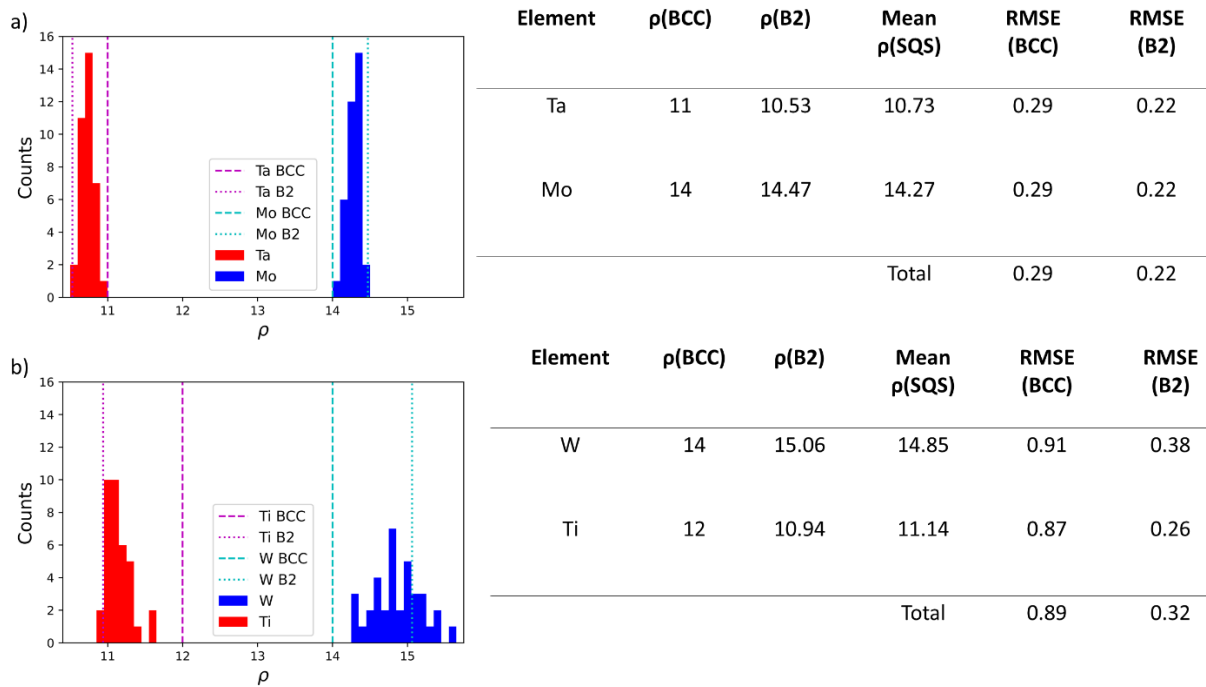

Figure 4) Histograms showing distributions of Bader charges for TaMo (a) and WTi (b). The Bader charge (ρ) of constituent elements as calculated in pure elemental BCC (dashed lines) and B2 Unit cells (dotted lines) are also shown for reference. Descriptive statistics are shown to the right of the histograms including the amount of error between BCC/B2 Bader charges and those observed in the corresponding SQS.

In addition to producing accurate lattice parameter predictions, the proposed model only requires the composition of a given alloy and the B2 lattice parameters in Table 1 as input without needing any further DFT calculations allowing efficient high throughput evaluation of lattice parameter. No fitting parameters are necessary in the proposed model eliminating the possibility that the results observed resulted from overfitting. It is important to note that Chromium was deliberately excluded from this study since it would introduce the need to account for magnetism in supercell relaxation which would have significantly increased the computational resources needed to collect data. As a result, alloys with chromium can not be evaluated in the present model. While the approach used in the present work could similarly investigate the effect of magnetism on bond lengths using B2 unit cells, DFT relaxation of large supercells to validate the model would be costly. Some error still remains in the proposed model which could potentially be explained by the nature of the B2 calculations used in the model which only accounts for first nearest neighbors which are assumed to all be equal and does not account for more distant interactions between a given bond and further atoms which could be an element different from the two nearest neighbors being considered. It was also noticed that alloys in which both Vegard's Law and the present work underestimated the lattice parameter had VECs lower than 4.5 suggesting another mechanism beyond charge transfer between dissimilar elements is involved. Another limitation is that the bond counting model used assumes truly random mixing in the alloys even though the presence of chemical short-range ordering (CSRO) has been proposed from simulation work on RHEAs[24] and observed experimentally in FCC HEAs[25,26]. While the model presented in this work is only applicable to BCC MPEAs. The general approach could also be applied to other crystal systems by using the bond counting model in concert with DFT relaxation of relevant ordered unit cells to account for charge transfer. For example, FCC or HCP alloys could be studied by performing DFT relaxations on $L1_0/L1_2$ or $DO_{19}$ ordered intermetallic unit cells respectively and using the values acquired from those calculations with a bond counting model.

      To summarize, the present study used DFT relaxations of B2 intermetallic compounds to obtain bond lengths between dissimilar atoms which was then combined with a bond counting model to produce a model to estimate the lattice parameters of Refractory BCC MPEAS. The model was tested using a data set of SQS supercells modeling various random solid solution MPEAs. The proposed model produced lattice parameters with superior accuracy to Vegard's Law without the need for additional DFT calculations or fitting parameters. The proposed model showed particularly impressive improvement over Vegard's Law for MPEAs containing elements with differing numbers of valence electrons and electronegativity since the DFT calculated bond lengths are able to account for changes in atomic volume due to charge transfer between dissimilar atoms. The proposed model is computationally lean allowing it to be used in high throughput combinatorial screening workflows. Further, the approach used of combining ordered intermetallic DFT calculations with a bond counting model has potential to be applied to materials systems beyond BCC HEAS composed of the elements studied in this work.

# References

1. Dalven, R. Empirical Relation between Energy Gap and Lattice Constant in Cubic Semiconductors. *Physical Review B* **8**, (1973).

2. Jiang, S. *et al.* Ultrastrong steel via minimal lattice misfit and high-density nanoprecipitation. *Nature* **544**, 460–464 (2017).

3. Gao, Y., Liu, Q. & Xu, B. Lattice Mismatch Dominant Yet Mechanically Tunable Thermal Conductivity in Bilayer Heterostructures. *ACS Nano* **10**, 5431–5439 (2016).

4. Shen, W. & Hertz, J. L. Ionic conductivity of YSZ/CZO multilayers with variable lattice mismatch. *Journal of Materials Chemistry A* **3**, 2378–2386 (2015).

5. Varvenne, C., Leyson, G. P. M., Ghazisaeidi, M. & Curtin, W. A. Solute strengthening in random alloys. *Acta Materialia* vol. 124 660–683 (2017).

6. Tyzack, C. & Raynor, G. v. The Lattice Spacings of Lead-rich Substitutional Solid Solutions. *Acta Crystallia* **7**, 505–510 (1954).

7. Raynor, G. V. The lattice spacings of the primary solid solutions of silver, cadmium and indium in magnesium. *Proceedings of the Royal Society of London. Series A. Mathematical and Physical Sciences* **174**, 457–471 (1940).

8. Hume-Rothery, W., Lewin, G. F. & Reynolds, P. W. The Lattice Spacings of Certain Primary Solid Solutions in Silver and Copper. *Proceedings of the Royal Society of London. Series A-Mathematical and Physical Sciences* **157**, 167–183 (1936).

9. Lee, J. A. & Raynor, G. v. The Lattice Spacings of Binary Tin-Rich Alloys. *Proc. Phys. Soc. B* **67**, 737–749 (1954).

10. Axon, H., Phil, D. & F, H.-R. The lattice spacings of solid solutions of different elements in aluminium. *Proceedings of the Royal Society of London. Series A. Mathematical and Physical Sciences* **193**, 1–24 (1948).

11. Vegard, L. Die Konstitution der Mischkistalle und die Raumfullung der Atome. *Zeitschrift fur Physik* **5**, 17–26 (1921).

12. Magnaterra, A. & Mezzetti, F. Charge transfer in binary alloys. *Nuov Cim B* **6**, 206–213 (1971).

13. Meng, F. *et al.* Charge transfer effect on local lattice distortion in a HfNbTiZr high entropy alloy. *Scripta Materialia* **203**, (2021).

14. Tong, Y. *et al.* Severe local lattice distortion in Zr- and/or Hf-containing refractory multi-principal element alloys. *Acta Materialia* **183**, 172–181 (2020).

15. Zunger, A., Wei, S.-H., Ferreira, L. G. & Bernard, J. E. *PHYSICAL REVIEW LETTERS Special Quasirandom Structures*. vol. 65 (1990).

16. Shu, S., Wells, P. B., Almirall, N., Odette, G. R. & Morgan, D. D. Thermodynamics and kinetics of core-shell versus appendage co-precipitation morphologies: An example in the Fe-Cu-Mn-Ni-Si system. *Acta Materialia* **157**, 298–306 (2018).

17. Clouet, E., Nastar, M. & Sigli, C. Nucleation of Al3Zr and Al3Sc in aluminum alloys: From kinetic Monte Carlo simulations to classical theory. *Physical Review B - Condensed Matter and Materials Physics* **69**, (2004).

18. Hu, Y. J., Sundar, A., Ogata, S. & Qi, L. Screening of generalized stacking fault energies, surface energies and intrinsic ductile potency of refractory multicomponent alloys. *Acta Materialia* **210**, (2021).

19. Blochl, P. E. *Projector augmented-+rave method*. PHYSICAL REVIEW B VOLUME vol. 50.

20. Perdew, J. P., Burke, K. & Ernzerhof, M. *Generalized Gradient Approximation Made Simple*. (1996).

21. Kresse, G. & Furthmü, J. *Efficient iterative schemes for ab initio total-energy calculations using a plane-wave basis set*. (1996).

22. van de Walle, A. *et al.* Efficient stochastic generation of special quasirandom structures. *Calphad: Computer Coupling of Phase Diagrams and Thermochemistry* **42**, 13–18 (2013).

23. Tang, W., Sanville, E. & Henkelman, G. A grid-based Bader analysis algorithm without lattice bias. *Journal of Physics Condensed Matter* **21**, (2009).

24. Fernández-Caballero, A., Wróbel, J. S., Mummery, P. M. & Nguyen-Manh, D. Short-Range Order in High Entropy Alloys: Theoretical Formulation and Application to Mo-Nb-Ta-V-W System. *Journal of Phase Equilibria and Diffusion* **38**, 391–403 (2017).

25. Chen, X. *et al.* Direct observation of chemical short-range order in a medium-entropy alloy. *Nature* **592**, 712–716 (2021).

26. Zhang, F. X. *et al.* Local Structure and Short-Range Order in a NiCoCr Solid Solution Alloy. *Physical Review Letters* **118**, (2017).